\documentclass[published]{epl}
\usepackage{graphicx}
\newcommand{\beq}{\begin{equation}}
\newcommand{\eeq}{\end{equation}}
\newcommand{\beqa} {\begin{eqnarray}}
\newcommand{\eeqa} {\end{eqnarray}}
\newcommand{\reff}[1]{(\ref{#1})}
\newcommand{\derpar}[2]{\frac{\partial #1}{\partial #2}}
  \newcommand{\Bibitem}[1]{\bibitem{#1}}
\newcommand{\Label}[1]{\label{#1}}
\title{Generalized Casimir forces in non-equilibrium systems }
\author{R.~Brito\inst{1} \and
U.~Marini Bettolo Marconi \inst{2} \and
R.~Soto \inst{3}}
\shortauthor{R.~Brito \etal}

\institute{
\inst{1} Dept. de F\'{\i}sica Aplicada I and GISC,
Universidad Complutense, 28040 Madrid, Spain\\
\inst{2} Dipartimento di Fisica, Universit\`a di Camerino, 
Via Madonna delle Carceri, 62032 Camerino,
Italy\\
\inst{3} Departamento de F\'{\i}sica, FCFM, Universidad de Chile,
Casilla 487-3, Santiago, Chile.}

\pacs{05.40.-a}{Fluctuation phenomena, random processes, noise, and
Brownian motion}
\pacs{05.20.Jj}{Statistical mechanics of classical fluids}
\pacs{74.40.+k}{Fluctuations (noise, chaos, nonequilibrium
superconductivity, localization, etc.)}

\begin{document}

\maketitle

\begin{abstract}
In the present work we propose a method to determine fluctuation induced forces
in non equilibrium systems. These forces are the analogue of 
the well known Casimir forces, which were originally introduced in Quantum 
Field theory and later extended to the area of Critical Phenomena.
The procedure starts from the observation that many non equilibrium systems
exhibit long-range correlations and the associated structure factors 
diverge in the long wavelength limit. The introduction of external bodies
into such systems in general modifies the spectrum of these fluctuations
and leads to the appearance of a net force between these bodies. 
The mechanism is illustrated by means of a 
simple example:
a reaction diffusion equation with random noises. 
\end{abstract}

\section{Introduction}

In 1948 Casimir \cite{CasimirOriginal} noticed that, since in the vacuum the 
quantum electromagnetic field
fluctuates rather than vanishing as in the classical
case, it would cost some energy to introduce
a macroscopic body into it. Indeed, 
the presence of a macroscopic body limits the allowed wave-vectors and
therefore the energy density. 
Casimir derived a simple expression for the resulting force per 
unit area acting between two large conducting neutral plates  
facing each other at distance $x$. 
The result is: $F/A=-\hbar c\pi^2/(240 x^4)$ and the presence of $\hbar c$
comes from the quantal origin of the force. 
In the last twenty years, Casimir's idea has been extended to some 
equilibrium systems where fluctuations are of thermal origin,
rather than quantal and thus the force is proportional to $k_BT$ \cite{Kardar}.
The unifying feature of systems displaying thermal Casimir
forces is the presence of {\em long ranged} {\em spatial correlations}. 
The best known case are binary fluid mixtures at the critical point
\cite{FisherandDeGennes},
where the Casimir force between two plates of area $A$ and separation $x$
is attractive and proportional to  $ -k_BTA/x^{3}$.
The inverse power law character of the force is 
originated by the divergence of the
correlation length, $\xi$, as the critical point is approached.
The Ornstein-Zernike theory~\cite{Stanley} predicts long range
density fluctuations characterized by a structure factor
(defined as the Fourier transform of the density-density correlation function)
like:
\beq
\Label{Scritical}
S(\vect{k}) \propto \frac{1}{\xi^{-2}+k^2},
\eeq
where  $\xi \propto |T-T_c|^{-1/2}$. Then the structure factor behaves as
$S(k)\propto k^{-2}$ at the critical point and the free energy 
of the system scales as $k_BT A/x^{d-1}$, from which one derives the
above expression for the force.

Casimir forces also appear in equilibrium physical systems
where long range correlations 
{\it appear because a continuous symmetry is broken}
\cite{Forster}. 
For instance in liquid crystals, with a vector order
parameter, the structure factor for a smectic-A phase layered in the
$z$-direction is anisotropic:
\beq\Label{Sliquidcrystal} 
S(\vect{k})\propto
\frac{k_\perp^2}{k_z^2+\lambda^2 k_\perp^4}. 
\eeq 
When the $\vect{k}$  is parallel to the layers, i.e. $k_z=0$, the
structure factor
diverges as $k^{-2}$. Such smectic liquid crystals also exhibit a Casimir
force: $F\propto-k_BTA/x^{2}$ \cite{AjdariPRL}. Similar long ranged 
correlations appear also in nematic phases of liquid crystals, or
magnetic liquid crystals immersed in a magnetic field $H$
\cite{deGennesProst}, 
in superfluid films \cite{Ueno}, in solutions of ionic salts
or micellar systems \cite{MartinCasielles} among many others.

Although there seems to exist a relation between the presence of long range
correlations and Casimir forces, this relation has not been elucidated
when computing the force. In practice, Casimir forces are obtained 
by differentiating with respect to the typical distance
the  free energy change \cite{Kardar}
induced by the confinement of the fluctuating field.

The goal of this Letter is to establish a general relation between
correlation functions (or the associated structure factors)
and Casimir forces. We also provide a novel method to compute
the Casimir force, specially useful for non-equilibrium
systems where long range correlations
are ubiquitous~\cite{Garrido},  but 
arguments based on the free energy cannot be invoked.

In a previous work~\cite{CasimirPRL} we derived
the Casimir force between large intruders in a granular system, starting
from the structure factors. In that case  the
balance between the energy dissipation due to collisions and the energy
injection, renders the system statistically stationary, but out of
equilibrium. As a consequence it develops long range 
correlations, and the structure factors for density, velocity and
temperature, as well as the cross structure functions, decay as $k^{-2}$
\cite{NoijeErnstPago}.

There are a variety of other non-equilibrium systems for which the 
structure factors have been obtained. 
A major class is represented by systems under
{\em spatial gradients}, such as an  isothermal uniformly 
sheared fluid in the
$z$-direction (plane Couette flow) \cite{Machta,Lutsko}. 
It was found that the structure factor in the limit $k\to 0$
behaves as $S(\vect{k})\propto -\dot\gamma_0\frac{k_xk_z}{k^4}$,
where $\dot\gamma_0$ is the shear rate.
In a similar fashion, a binary fluid in a Rayleigh-Benard cell with
a temperature gradient $\nabla T$ below the onset of the
instability, shows an enhancement of the structure factors
proportional to $(\nabla T)^2/k^4$. Here the fluctuations are
measured in the $xy$-plane, orthogonal to the gradient of the
temperature~\cite{Li}. In the same Rayleigh-Benard cell, but heated
from above, Li et al. showed that the concentration fluctuations in a
polymer solution behave as $(\nabla c)^2/k^4$, where $\nabla c$ is
the concentration gradient created by the Soret effect. 
Pure concentration gradient were studied by Spohn \cite{Spohn} using
lattice gases, who showed that the structure factors behave like $(\nabla
c)^2/(k^2+k_0^2)$, with $k_0=\pi/(2L)$, being $L$ the system size.

Another class of non-equilibrium systems exhibiting long range correlations are
those lacking the {\em  detailed balance condition}. For instance,
it has been shown \cite{Grinstein,Pagonabarraga} that systems
described by the Langevin equation with conservative dynamics and
non-conservative noise have long range spatial and temporal
correlations. Analogously, non-equilibrium concentration fluctuations
in reaction diffusion systems can be long ranged under certain
conditions \cite{Gardiner} that will be described in the next section.

%%%%%%%%%%%%%%%%%%%%%%%%%%%%%%%%%%

\section{The mesoscopic model}
As an example of how a Casimir force arises in a
non-equilibrium system with long range correlations, we consider the
simple case of a reaction-diffusion system in three dimensions, 
where the fluctuating density $n$ around the homogeneous
reference density $n_0$ obeys the equation
\beq
\derpar{\phi}{t} = \nabla\cdot(D\nabla \phi+\vect{\xi}_1) - \lambda \phi
+\xi_2, \Label{eq.reacdif}
\eeq
where $\phi=n-n_0$ is the fluctuating field, $D$ is the diffusion
coefficient, and $\lambda>0$ is the relaxation rate. The terms
$\vect{\xi}_1$ and $\xi_2$ describe fluctuations in the
diffusive flux and in the reaction rate. They are assumed to have white
noise spectrum
\beqa
\langle \xi_{1i}(\vect{r},t) \xi_{1k}(\vect{r}',t') \rangle &=&
\Gamma_1 \delta_{i,k}\delta(\vect{r}-\vect{r}')\delta(t-t')\nonumber \\
\langle \xi_{2}(\vect{r},t) \xi_{2}(\vect{r}',t') \rangle &=&
\Gamma_2 \delta(\vect{r}-\vect{r}')\delta(t-t')
\eeqa
where $\Gamma_1$ and $\Gamma_2$ are the noise intensities.
This reaction-diffusion equation models, 
for instance, the set of chemical reactions:
$A\stackrel{k_1}{\longrightarrow}B$,
$A+B \stackrel{k_2}{\longrightarrow}2A$, that do not satisfy 
the detailed balance condition \cite{WOH}.  
In this case, $\phi$ is the density fluctuation of either
the species $A$ or $B$ relative to its average stationary density
and $\lambda= n_0k_2-k_1$. When $\lambda=0$ the reaction shows a critical point
at density $n_0 = k_1/k_2$.
We assume that the pressure in the system is a function of the local density $p(n)$.

Equation~\reff{eq.reacdif} not only models the chemical reaction described above, 
but governs many other physical systems, like liquid crystals \cite{Ajdari} or
superfluid films \cite{Kardar}. 
All these systems possess structure
factors similar to those mentioned in the Introduction and the results obtained 
in this Letter can be generalized to those systems. 

The solution of Eq. \reff{eq.reacdif} predicts that,
after an initial transient, 
the density $\phi$ is statistically homogeneous and stationary so that
in Fourier space we have:
\beq
\langle \phi_{\vect{k}} \rangle =0, \quad
\langle \phi_{\vect{k}} \phi_{\vect{q}} \rangle = V S({\vect{k}})
\delta_{{\vect{k}},-{\vect{q}}} 
\eeq
where the symbol $\langle
\cdot\rangle$ represents the average over the two noises $\vect{\xi}_1$ and
$\xi_2$. The structure factor $S(\vect{k})$  is given by (see  e.g.~Chap.~(8.3)
of \cite{Gardiner}):
\beq
S({\vect{k}}) = \frac{\Gamma_1 k^2+\Gamma_2}{2(D k^2+\lambda)}
= \frac{\Gamma_1}{2D} + \frac{\Gamma/2D}{k^2+k_0^2} \Label{eq:Sk}
\eeq
with $\Gamma=\Gamma_2-\Gamma_1\lambda/D$ and $k_0=\sqrt{\lambda/D}$.
The corresponding real space correlation reads
\beq
G({\vect{r}}) = \frac{\Gamma_1}{2D}
\delta(\vect{r}) + \frac{\Gamma}{2D} \frac{e^{-k_0 r}}{r}
\Label{eq:Yukawa}\eeq 
The second contribution, stemming  from the  $k$-dependent term in 
\reff{eq:Sk}, represents fluctuations with a
correlation length that depends on the reaction parameter, $\lambda$, and,
therefore,  are of macroscopic size. In particular, 
near the critical point
$\lambda\to0$, the correlation length diverges. If the reaction
satisfies the fluctuation-dissipation theorem \cite{Gardiner,Ojalvo}
then $\Gamma_1=2k_BTD$ and
$\Gamma_2=2k_BT\lambda$, where $T$ is the temperature,
implying that $\Gamma$ vanishes along with the
macroscopic correlations. 
On the contrary, in non-equilibrium
systems which violate the fluctuation-dissipation theorem,
$\Gamma$ does not vanish and macroscopic correlations are present.
The $\delta$-term in Eq.~\reff{eq:Yukawa}, 
coming from $\Gamma_1/2D$, describes the microscopic 
self-correlation of the particles that a mesoscopic model cannot resolve. 
Therefore, they can be eliminated. This corresponds to
subtracting  the asymptotic value of $S$ for large values of $k$.
From now on, we will consider the macroscopic part (or, equivalently, the
non-equilibrium part) of the structure factor
$ S^*(\vect{k})=S(\vect{k})-\lim_{q\to\infty}
S(q)=\Gamma/[2D(k^2+k_0^2)]$.
This is equivalent to suppress the vectorial noise $\vect{\xi}_1$ and keep
only a scalar noise $\xi$ with an intensity $\Gamma$.

We study now the effect of confining 
the system between two plates, parallel and
infinite in the $y$ and $z$-directions, located at $x=0$ and $x=L$, 
and calculate the force between the plates surrounded by a fluid 
described by Eq.~(\ref{eq.reacdif}). The force derives from 
the pressure, $p(n)$, exerted by the particles over the plates.
To proceed, we consider the system in a volume $L_x\times
L_y\times L_z$, periodic in all directions. In this volume we place two
plates at distance $L$ with non flux boundary conditions at them,
as natural for a reacting system.
The total volume $V$ results divided into two regions: 
Region I in between the plates of
volume $L\times L_y\times L_z$, and Region II outside the plates of volume
$(L_x-L)\times L_y\times L_z$.  The limit $L_x,L_y,L_z\to\infty$
will eventually be taken.

In order to perform the analysis in the two regions let us consider a case
of a general volume $V=X\times L_y\times L_z$, where $X=L$ for Region I and
$X=L_x-L$ for Region II. The density field is expanded, taking into account
the non flux boundary conditions on the $x$-direction, as
\beq
\phi(\vect{r},t) = V^{-1} \sum_{k_x}\sum_{k_y} \sum_{k_z}
\phi_{\vect{k}}(t) \cos(k_x x) e^{ik_y y}e^{ik_z z}
\label{eq.expansionn}
\eeq
where $k_x=\pi n_x/X$, $k_y=2\pi n_y/L_y$, $k_z=2\pi n_z/L_z$,
$n_x=0,1,2,\ldots$ and 
$n_y,n_z=\ldots,-1,0,1,\ldots$.
The noise $\xi$ is expanded in a similar way with
\beq
\langle \xi_{\vect{k}}(t) \xi_{\vect{q}}(t') \rangle =
\gamma_{k_x} V \Gamma \hat\delta_{\vect{k},\vect{q}} \delta(t-t')
\eeq
where $\hat\delta_{\vect{k},\vect{q}}=
\delta_{k_x,q_x}\delta_{k_y,-q_y}\delta_{k_z,-q_z}$ is a modified 3D
Kronecker delta.
Moreover the factor $\gamma_{k_x}$ ($\gamma_{k_x}=1/2$ if $k_x=0$ and  
$\gamma_{k_x}=1$ otherwise)
appears because of the non-flux boundary condition in the $x$-direction.
Replacing these expansions in \reff{eq.reacdif} it is found that
\beq
\langle \phi_{\vect{k}} \rangle =0, \quad
\langle \phi_{\vect{k}} \phi_{\vect{q}} \rangle =
\gamma_{k_x}\hat\delta_{\vect{k},\vect{q}}
V  S^{*}(\vect{k})
\eeq
with the same structure factor $S^*(\vect{k})$ as in the homogeneous case.
Finally, the density field fluctuations in real space are given by: 
\beqa 
\langle \phi(\vect{r}) \rangle =0, \quad
\langle \phi(\vect{r})^2 \rangle
= \frac{\Gamma}{2Dk_0^2}V^{-1}
{\sum_{\vect{q}}}^\prime \frac{1}{q^2+1} \cos(q_x k_0 x)^2.
\Label{eq:phiq}
\eeqa
where $\vect{q}=\vect{k}/k_0$  and the prime in the sum means that the term
$q_x=0$ has a factor 1/2.

The sum in Eq.~(\ref{eq:phiq}) contains an ultraviolet divergence 
($\vect{q}\to\infty$). Therefore, in order to perform the summation
a regularization prescription is needed.
We introduce a regularizing kernel  in Eq.~\reff{eq:phiq} of the form
$1/(1+\epsilon^2 q^2)$, that equals 1 for $\epsilon\to 0$, 
limit that will be taken
at the end of the calculations. The election of a rational function
instead of an exponential one is made to keep the calculations as simple 
as possible. 
The technique of a regularizing kernel is equivalent to
imposing a cutoff in the $q$-vectors of the order of $q_c\sim\epsilon^{-1}$ 
or to the {\it zeta-function} regularization method \cite{Ajdari}.

Next, we take the limit $L_y,L_z\to\infty$  allowing us to replace the
sums on $q_y$ and $q_z$ by
integrals that can be carried out, with the result:
\beq
\langle \phi({\vect{r}})^2 \rangle = \frac{\Gamma}{8\pi D X }
\frac{1}{1-\epsilon^2}
{\sum_{q_x}}^\prime
\log\left(
\frac{1+\epsilon^2q_x^2}{\epsilon^2(1+q_x^2)}
\right) \cos(q_xk_0 x)^2. \label{eq:perfil}
\eeq

\begin{figure}
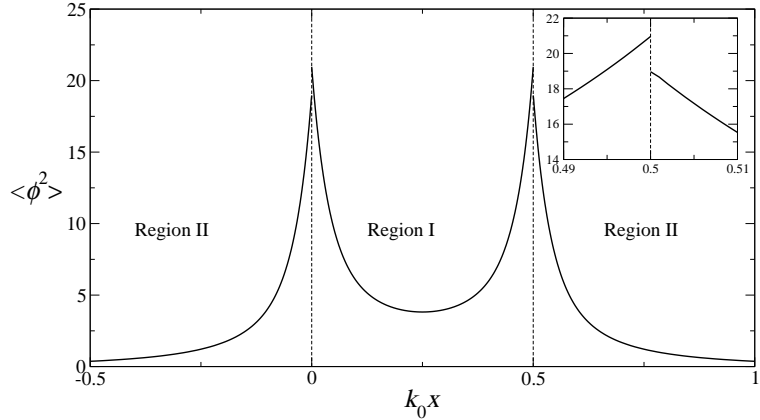

\onefigure[width=0.7\columnwidth]{CasimirFig1.eps}
\caption{Density fluctuations as a function of the dimensionless position $k_0x$,
for $\epsilon=0.05$ and $k_0L=1/2$. The vertical scale is in arbitrary units and the
asymptotic value of the density fluctuations has been subtracted. The inset
shows the density fluctuations jump at the plate. }
\label{Fig:1}
\end{figure}

Figure \ref{Fig:1} shows the density fluctuations \reff{eq:perfil} 
in the $x$-direction when the plates are located at $x=0$ and 
$k_0L=1/2$, $\epsilon=0.05$ and $L_x\to\infty$.
The set of $q_x$-vectors entering in the sum of 
Eq.~\reff{eq:perfil} for Regions I and II are different.  
In Region I, the allowed $q_x$ vectors are quantized 
as: $q_x=\pi n_x /(k_0L)$, while in
Region II they form a continuum in the limit $L_x \to\infty$.
This difference produces a jump of $\langle\phi^2\rangle$ at the plate, that is
shown in the inset.

\section{Casimir forces}
As shown in Fig. \ref{Fig:1}, the density fluctuations in
a confined system depend on the position but also on the system size, and
therefore  are different in the region in between the plates and the
region outside them. If the
pressure is a function of the local density field $p(n)$, these
differences in the density fluctuations create an unbalance
in pressure between both
sides of each plate, and consequently a net force.
To calculate the force, we expand the
local pressure around the reference density $n_0$, and take the
statistical average, finding that
\beq
\langle p(x) \rangle = p(n_0) +
\frac{1}{2}\left.\derpar{^2p}{n^2}\right|_{n_0}
\langle \phi(x)^2 \rangle.
\eeq
The net force acting on the plate 
located at $x=L$ is the difference between the
pressure inside $p(x\to L^-)$ 
and the pressure outside $p(x\to L^+)$, that come from 
density fluctuations in Regions I and II respectively. 
Then the net effective force between the plates by unit area is twice the
pressure difference in one of the plates:
\beq
F/A= \frac{\Gamma(\partial^2p/\partial n^2)}{16\pi D(1-\epsilon^2)}
\left[\frac{1}{L}
\sum_{q_x=-\infty}^\infty \log\left(
\frac{1+\epsilon^2q_x^2}{\epsilon^2(1+q_x^2)} \right)
-\frac{k_0}{\pi}
\int_{-\infty}^{\infty} dq_x
\log\left(
\frac{1+\epsilon^2q_x^2}{\epsilon^2(1+q_x^2)}
\right)\right]
\Label{eq:Forceeps}\eeq
The integral in this equation is simply $2\pi(\epsilon-1)/\epsilon$, and
Eq.(1.431,2) of \cite{Gradstheyn}, allows us to evaluate the sum. The
diverging terms ${\cal O}(\epsilon^{-1})$ cancel and the result for the
force, in the limit of a vanishing cutoff $\epsilon$, is surprisingly
simple:
\beq
F/A=F_0\left(1-\frac{\log(2\sinh l)}{l}\right)
\Label{eq:Force}\eeq
where $l=k_0L_x$ and $F_0=\Gamma k_0(\partial^2p/\partial n^2) /(8\pi D)$.
Let us note that the final expression of the Casimir force, is a {\em universal}
function of the reduced distance, $l=k_0L_x$. Moreover, there is no dependence
on the cutoff length, as the two divergences in the cutoff, one stemming
from
the discrete sum and another from the integral, exactly cancel each other.
The regularizing kernel, a technique well known in the field of
Casimir forces \cite{renormalization},
has allowed us to obtain a finite result as a difference
of two diverging quantities. 

The analysis of the Casimir force, Eq. (\ref{eq:Force}) can be 
performed in
the limits of far plates ($l\gg1$) or near ones ($l\ll 1$). 
In the first case, ($l\gg1$),
implies that the distance $L_x\gg k_0^{-1}$ and therefore the 
plates are outside the
correlation length, $k_0^{-1}$ (see Eq. (\ref{eq:Yukawa})). 
Then, one expects a very
fast decay of the Casimir forces. In the opposite limit ($l\ll1$), when the
plates are well inside the correlation length, the force
is much stronger. These forces are
\beq
F^{far}/A=F_0 \frac{e^{-2l}}{l}, \quad F^{near}/A=-F_0 \frac{\log l}{l}.
\Label{eq:farnear}
\eeq

\begin{figure}[h]
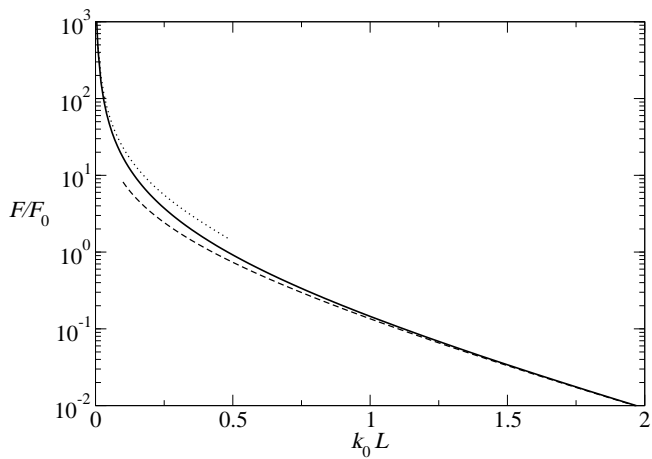

\onefigure[width=0.6\columnwidth]{CasimirFig2.eps}
\caption{Dimensionless force as a function of the dimensionless distance
$l$, together with
the asymptotic form for $l\gg 1$ (dashed line) and $l\ll 1$ (dotted line).
}\label{Fig:2}
\end{figure}

These results are presented in Fig. \ref{Fig:2} where we plot the exact
force Eq. \reff{eq:Force} as a function of the dimensionless distance $l$,
together with the far and near plate approximations.

\section{Summary and Conclusions} We have calculated the fluctuation
induced Casimir force between two parallel plates produced by a
scalar field whose evolution is given by a
reaction-diffusion equation. This system is out of
equilibrium because the noise terms do not satisfy the
fluctuation-dissipation relation, developing
long range correlations. From these correlations
one can calculate the pressure that the fluctuating field 
exerts on the two parallel plates.

The Casimir force obtained in this way depends on a macroscopic parameter
$k_0$ that defines a characteristic length as $k_0^{-1}$, which is a
combination of the diffusion constant and reaction rates, similar to the
cases described in other physical systems \cite{Spohn,WOH}. The calculated
force exhibits a crossover at plate
separation of the order of $k_0^{-1}$, distance at which the
correlation have decayed (see Eq. (\ref{eq:Yukawa})). For short
separations, the correlations play an important role and the 
force shows a strong distance dependence, while for $L\gg k_0^{-1}$ the 
density correlations have decayed and the force vanishes 
exponentially.

Finally the force is repulsive if the pressure is such that 
$\partial^2 p/\partial n^2>0$, as it is usually the case for molecular
fluids far form the liquid-gas transition. However,
the sign might change for different geometries of the obstacles (for
instance plate-sphere, or two concentric spheres \cite{Boyer}). 

To conclude, we have developed a method to calculate Casimir forces
starting from the structure factors. This method is valid for both
equilibrium and non equilibrium systems, where the usual derivation of
Casimir forces based on free energy calculations is not applicable. The
necessary ingredients to obtain a Casimir force are a fluctuating field
with correlations that extend over a macroscopic range,  and a confinement
of
the fluctuation spectrum induced by the presence of the plates or other
obstacles.

{\it Acknowledgements.}
The authors thank to J.M.~Ortiz for useful comments. 
R.B. acknowledges the hospitality of the Dpt.~de F\'{\i}sica of
Universidad de Chile. 
R.B. is supported by Secretar\'{\i}a de Estado de Educaci\'on
y Universidades (Spain) and the Projects No.
FIS04-271 (Spain) and No. UCM PR27/05-13923-BSCH. 
U.M.B.M. acknowledges a grant COFIN-MIUR 2005, 2005027808.
R.S. acknowledges the hospitality of the Dpt.~de F\'{\i}sica Aplicada I of UCM. 
R.S. is supported by {\em Fondecyt} research grants 1030993 
and 1061112 and {\em Fondap} grant 11980002. 

%%%%%%%%%%%%%%%%%%%%%%%%%%%%%%%%%%%%%%%%

\end{document}